\documentclass[11pt,twoside]{article}
\usepackage{macroaaa-eng}
\usepackage{psfig}
\usepackage{graphicx}

\usepackage[T1]{fontenc} 

\usepackage{latexsym}
\usepackage{verbatim}

\newcommand{\mo}    {M$_{\odot}$}

\newcommand{\gr}    {$^{\circ}$}
\newcommand{\kms}    {km~s$^{-1}$}
\newcommand{\ltsima} {$\; \buildrel < \over \sim \;$}
\newcommand{\simlt}  {\lower.5ex\hbox{\ltsima}}         

\begin{document}

\vskip 1.0cm
\markboth{J.~M.~Paredes}{Black Holes in the Galaxy}
\pagestyle{myheadings}

\vspace*{0.5cm}
\title{Black Holes in the Galaxy}

\author{Josep~M.~Paredes}
\affil{Departament d'Astronomia i Meteorologia and Institut de Ci\`encies del Cosmos (ICC), Universitat de Barcelona (UB/IEEC), Mart\'{\i} i 
Franqu\`es 1,
08028 Barcelona, Spain\\E-mail: jmparedes@ub.edu}

\begin{abstract} The most compelling evidences for the existence of stellar-mass black holes have been obtained through observations of X-ray binary systems. The application of classical methods and the development of new techniques have allowed us to increase the number of stellar-mass black holes known. I summarize here the observational signatures of the black holes, such as the mass determination, the event horizon and the spin. I also present some observational results on the Galactic centre black hole.
\end{abstract}

\section{Introduction}

The contribution presented here is part of a course on Compact Objects and their Emission given in the First La Plata International School on Astronomy and Geophysics. The contribution is focused on the observational evidences of stellar mass black holes in our Galaxy and the super-massive black hole in its centre. The four Lectures that were given about the black holes in the Galaxy have been distributed here in five Sections.

In Section 2 a brief account about how a neutron star or a black hole are formed is presented, a brief black hole history is also introduced and some concepts such as the Schwarzchild radius, the event horizon and Kerr black hole are given. As the X-ray binaries play a fundamental role for the study of stellar mass black holes, Section 3 is devoted to introduce the X-ray binaries and explain some of their most important characteristics. In Section 4, the first black hole candidates are introduced, how were they discovered and how many of them are currently known. Section 5 is devoted specifically to introduce the observational signatures of the black holes. In particular, some methods to estimate the mass, to find possible evidences for the event horizon and to measure the black hole spin are explained. Finally, Section 6 is devoted to the super-massive black hole in the centre of our Galaxy.

\section{Physical background}

\subsection{Dead stars}

During their life stars evolve through different states that are linked to the nuclear burning. 
When the core of the star runs out of nuclear fuel, it collapses until some other form of pressure support enables a new equilibrium configuration to be attained. The possible equilibrium configurations which can exist
when the star collapses are: White Dwarfs (WD), Neutron Stars (NS), Black Holes (BH). According the initial star mass ($M_{\rm init}$), there will be different types of collapse and different supernova explosions. When $M_{\rm init}$ < 8\,\mo, there is no SN explosion, only contraction, and a WD  is formed.
If the WD accretes mass it will explode as a type Ia SN. When $M_{\rm init}$ > 8\,\mo, thermonuclear SN explosions of different types are produced (types Ib, Ic and no H lines; Types II and H lines). As a result, a NS or a BH is formed. 

In WD and NS, the pressure which holds up the star is the quantum mechanical pressure associated with the fact that electrons, protons and neutrons are fermions (only one particle is allowed to occupy any one quantum mechanical state). WD are held up by electron degeneracy pressure, and their mass is below the Chandrasekhar limiting mass $M_{\rm Ch}$ = 1.46\,\mo. With increasing density, the degenerate electron gas becomes relativistic and, when the total energy of the electron exceeds the mass difference between the neutron and the proton, the inverse $\beta$-decay process can convert protons into neutrons. It is the degeneracy pressure of this neutron gas which prevents collapse under gravity and results in the formation of a NS.

\begin{figure}  
\begin{center}
\hspace{0.25cm}
   \psfig{figure=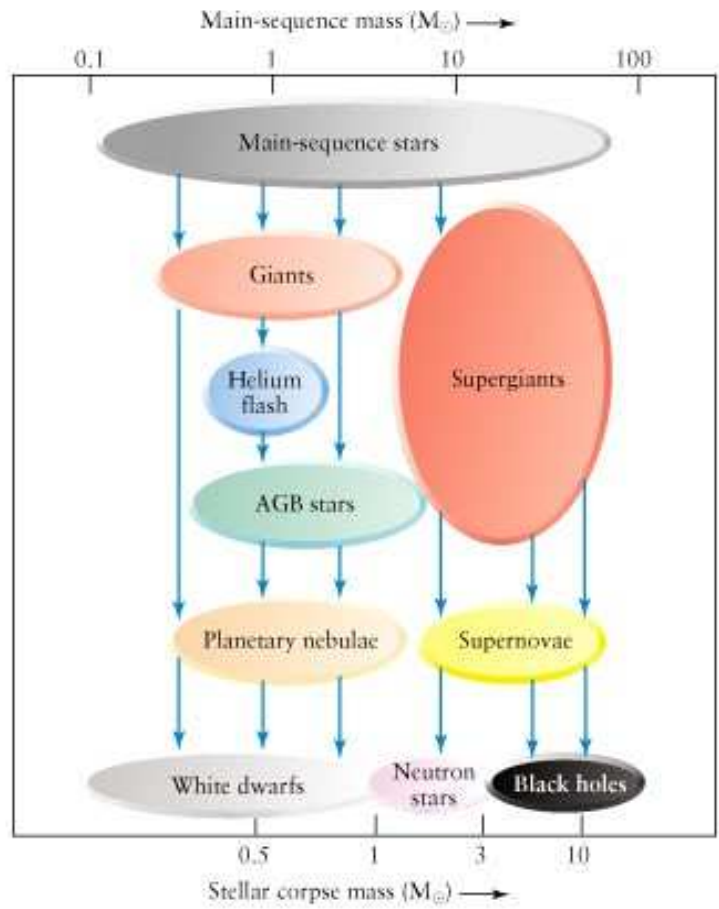,width=10.cm}
\caption{Relation between the original mass and the final mass of a star (from Comins-Kaufmann 1991).}
\label{mass}
\end{center}
\end{figure}

The same physics for the WD is responsible for providing pressure support for the NS, the only difference being that the neutrons are about 2000 times more massive than the electrons, and consequently degeneracy sets in at a correspondingly higher density.
NS, in which neutron degeneracy pressure is responsible for the pressure support, can have a mass given by the Tolman-Oppenheimer-Volkoff mass limit. This mass is not analytically well fixed because of 
its dependence on the equations of state for
nuclear matter. In any case, most equations of state do not allow the neutron degeneracy pressure to support more than 3\,\mo, indicating that dead stars more massive than 3\,\mo must be BHs. Figure~\ref{mass} shows an scheme of the relation between the original mass and the final mass of a dead star. 

\subsection{Brief Black Hole History}

John Michell first described the concept of a black hole in a paper that appeared in Philosophical Transactions of the Royal Society of London in 1783. Michell discussed the possibility of a star with a gravitational force so strong that not even light could escape from it, thus preventing astronomers from observing such phenomena. 
Pierre-Simon Laplace mentioned the possibility of such stars in the first few editions of his book Exposition du syst\`eme du Monde (1796), although he failed to include it in later editions because the idea of a star with a gravitational force that could overpower light did not comply with the wave theory of light, which was generally accepted at that time. This theory seemed to suggest that light could not be affected by gravity, a concept that is key to the theories of Michell and Laplace. 
In the early 1800's experiments on optical interference led to the predominance of the wave theory of light and the end of the corpuscular theory. Since light waves were thought to be unaffected by gravitation, interest in the hypothetical "dark stars" ceased. 

In 1905 Albert Einstein published his Special Theory of Relativity and in 1915 his General Theory of Relativity (GR). The GR was a new theory of gravitation and one of its fundamental predictions was the effect of gravity on light. Matter causes space-time to curve and therefore the paths followed by light rays or matter are determined by the curvature of the space-time. This allowed for a modern scientific proof of Michell's hypothesis. 
   
A short time after Einstein published his GR, the German physicist Karl Schwarzchild wrote a paper describing a structure called a singularity. He
argued that matter could theoretically be drawn into a point with virtually no volume and a infinite density. He called this object a point mass, later dubbed a singularity. In addition, Schwarzchild determined that there is a definite boundary around a singularity called the event horizon. 

In 1928, through his research on WD, Subrahmanyan Chandrasekhar hypothesized that a dying star of a certain mass might form a point with enough gravitational pull to trap light. 

In 1939, American physicist Robert Oppenheimer developed a possible explanation for the nature of these points of infinite density. Oppenheimer theorized that the gravitational pull of a star with infinite density would cause light rays to deviate from their path and bend towards the star. Eventually, the gravity of the star would become so great that the light would become trapped by it and would be unable to escape, preventing one from observing it. At this point the star is a BH.
The American scientist John Wheeler coined the term "black hole" in 1969, although from a non astrophysical point of view, the expression "black hole" was coined before in Calcutta to describe a small dungeon  where troops of the Nawab of Bengal held British prisoners of war after the capture of Fort William on June 20, 1756.

\subsection{Schwarzchild radius, Event horizon, Kerr BH}

Performing a classical calculation, the escape velocity from the surface of a star of mass $M$ and radius $r$ is

\begin{equation}
\frac{1}{2}mv_{\rm e}^2=\frac{GMm}{r}
\qquad
\rightarrow
\qquad
v_{\rm e} = \sqrt{2GM/r}
\end{equation}

and setting $v_{\rm e} = c$, the radius of such star would be $r = 2GM/c^{2}$.
As we will see later, this is just the expression for the Schwarzchild radius of a BH of mass $M$, and the spherical surface of radius $r$ plays the role of the surface of the BH. The escape velocity from the Earth is 11\,km s$^{-1}$, from the Sun is 617\,km s$^{-1}$, from a NS (2\,\mo and radius 10\,km) is 230\,000\,km s$^{-1}$ and from a BH should be greater than $c$.

Some months after the Einstein's definitive formulation of the GR in 1915,
Schwarzschild discovered the solution for the metric of space-time about a 
point mass $M$:

\begin{equation}\label{eq:metric}
{\rm d}s^2=\Big( 1-\frac{2GM}{rc^2}\Big){\rm d}t^2-\frac{1}{c^2}\Bigg[\frac{{\rm d}r^2}{\Big(1-\frac{2GM}{rc^2}\Big)}+r^2({\rm d}\theta^2+{\sin}^2 \theta {\rm d}\phi^2)\Bigg]
\end{equation}

This metric, known as the Schwarzschild metric, has the same meaning as 
the interval d$s^2$ in special relativity, which is known as the Minkowski metric:

\begin{equation}
{\rm d}s^2={\rm d}t^2-\frac{1}{c^2}\Big[{\rm d}r^2+r^2({\rm d}\theta^2+{\sin}^2 \theta {\rm d}\phi^2)\Big]
\end{equation}

In the limit of large distances from the point mass, $r \to \infty$, the two metrics become the same. They are, however, very different for small values of $r$, reflecting the influence of the mass $M$ upon the geometry of space-time. 
In the case of the Schwarzschild metric, the time interval between events
according to an observer who is stationary in S and located at the point $r$ is d$t'= $d$t (1 - 2GM / r c^2)^{1/2}$.

This notation makes it clear how the time interval d$t^\prime$ depends upon the
gravitational potential in which the observer is located.
d$t^\prime$ only reduces to d$t$ in the limit of very large distances from the origin, $r \to \infty$, at which the gravitational potential goes to zero.

{\it Redshift of electromagnetic waves}.
If the time interval $\Delta t^\prime$ corresponds to the period of the waves emitted  at the 
point $r$, the observed period of the wave at infinity $\Delta t$ is given by

\begin{equation}
\Delta t^\prime=\Bigg(1-\frac{2GM}{rc^2}\Bigg)^{\frac{1}{2}}\Delta t
\end{equation}

Therefore, the emitted and observed frequencies, $\nu_{e}$ and $\nu_{\infty}$ , respectively, are related by 

\begin{equation}
\nu_{\infty}=\nu_{e}\Bigg(1-\frac{2GM}{rc^2}\Bigg)^{\frac{1}{2}}
\end{equation}

This expression is the general relativistic result corresponding to Michell's 
insight.
If the radiation is emitted from radial coordinate $r = 2GM / c^2$, the frequency
of any wave is redshifted to zero frequency, and  no information can reach infinity from radii $r  \le r_{\rm Sch} = 2GM / c^2$. Here $r_{\rm Sch}$ is the {\it Schwarzschild radius}. Since, according to GR, no radiation can escape from within this radius, the surface $r = r_{\rm Sch}$ is black.

The Schwarzschild radius defines the {\it event horizon}. Not even light can escape,
once it has crossed the event horizon. It can be written as $r_{\rm Sch} = 3 \Big(\frac{M}{M_{\odot}}\Big)$ km.

Another important aspect of the Schwarzschild metric is the dynamics of test masses
in the gravitational field of the point mass. In a Newtonian treatment, from the conservation of energy we obtain

\begin{equation}
\frac{1}{2}mv^2-\frac{GMm}{r}=\frac{1}{2}mv_{\infty}^2 
\qquad
\rightarrow
\qquad
\dot{r}^2+(r\dot{\theta})^2-\frac{2GM}{r}=v_{\infty}^2
\end{equation}  

and from the conservation of angular momentum, $m\dot{\theta}r^2= {\rm constant}$. Introducing the specific angular momentum of the particle
(angular momentum per unit mass) $h=\dot{\theta}r^2$, we can write

\begin{equation}
\dot{r}^2+\frac{h^2}{r^2}-\frac{2GM}{r}=\dot{r}_{\infty}^2
\end{equation}  
 
As long as $h \ne 0$, the particle cannot reach $r = 0$ because 
the energy term associated with the centrifugal force, $h^2/r^2$, becomes greater than
the gravitational potential energy $2GM/r$ for small enough values of $r$.

In a General Relativistic treatment the dynamics of the test mass is given by

\begin{equation}
\dot{r}^2+\frac{h^2}{r^2}-\frac{2GM}{r}-\frac{2GMh^2}{r^3c^2}=(A^2-1)c^2
\end{equation}  

There are two main differences when comparing  this relativistic equation with the Newtonian equation. There is an extra term and constants and variables have different meaning in GR ($r$ angular diameter distance, $\dot{r}$ means d$r$/d$s$, where $s$ is proper time,
$h$ and $(A^2-1)$ are constants equivalent to $h$ and $\dot{r}_{\infty}^2$). The most important difference is the additional term $\frac{2GMh^2}{r^3c^2}$, which has the effect of enhancing the attractive force of gravity, even when the particle
has a finite specific angular momentum $h$ (the greater the value of $h$, the greater the enhancement). This result can be understood by recalling that the kinetic energy associated
with the rotational motion around the point mass contributes to the inertial mass of a
test particle and thus enhances the gravitational force upon it. From the analysis of the relativistic
equation, some interesting results can be found: \\
1) For sufficiently small values of $r$, the general relativistic term$\frac{2GMh^2}{r^3c^2}$ becomes greater than the centrifugal potential term, implying that this purely general relativistic term increases the strength of gravity close enough to $r = 0$. It can be shown easily that if the specific angular momentum of the particle $h \le 2r_{\rm Sch}c$ it will inevitably fall in to $r = 0$. \\
2) There is a last stable circular orbit, of radius $r = 3 r_{\rm Sch}$, around the point mass. There are no stable circular orbits with smaller radii than this value
because the particles would spiral rapidly into $r = 0$. This is why the BH is called hole, matter inevitably collapses in to $r = 0$ if
it comes too close to the point mass. \\
3) From the metric given in Eq. \ref{eq:metric}, it appears that there is a singularity at $r_{\rm Sch}$. It can be shown that this is not a physical singularity. However, at $r = 0$, there is a real physical singularity, and according to the GR, the
infalling matter collapses to a singular point.However, these Schwarzschild singularities are unobservable because no
information can arrive to the external observer from within $r_{\rm Sch}$. For all practical purposes, the BH may be considered to have a black spherical
surface at $r_{\rm Sch}$. From the classical point of view, physics breaks down at $r = 0$.

{\it Kerr black holes}. In 1962, Kerr discovered the general solution for a BH with angular momentum $J$.
It has been shown that isolated BHs can be completely characterized by only three properties:
mass $M$, charge $Q$ and angular momentum $J$. The rotating BHs (Kerr BHs) are relevant to many aspects of high energy astrophysics.
The Kerr metric in Boyer-Landquist coordinates is given by:

\begin{eqnarray}
{\rm d}s^2=\Bigg(1-\frac{2GMr}{\rho c^2}\Bigg){\rm d}t^2-\frac{1}{c^2}\Bigg[\frac{4GM\,r\,a\,{\rm sin}^2\theta}{\rho c}{\rm d}t\,{\rm d}\phi+\frac{\rho}{\Delta}{\rm d}r^2+\rho\,{\rm d}\theta^2+
\nonumber\\
{}+\Bigg(r^2+a^2+\frac{2GM\,r\,a^2\,{\rm sin}^2\theta}{\rho c^2}\Bigg){\rm sin}^2\theta\,{\rm d}\phi^2\Bigg]
\end{eqnarray}

where $a=J/Mc$ is the angular momentum of the BH per unit mass, $\Delta=r^2-(2GMr/c^2)+a^2$ and $\rho=r^2+a^2{\rm cos}^2\theta$.
If the BH is non-rotating, $J = a = 0$ and the Kerr metric reduces to the 
standard Schwarzschild metric. Just as in the case of Schwarschild metric,  the metric coefficient of d$r^2$ becomes 
singular at a certain radial distance, when  $\Delta = 0$. This radius correspond to the surface of infinite redshift or the horizon of the 
rotating BH, and is given by the solution of $\Delta=0$. Taking the larger of the two roots, the horizon occurs at radius

\begin{equation}
r_{\rm +}=\frac{GM}{c^2}+\Bigg[\Bigg(\frac{GM}{c^2}\Bigg)^2-\Bigg(\frac{J}{Mc}\Bigg)^2\Bigg]^{\frac{1}{2}}
\end{equation}

This spherical surface has exactly the same properties as the Schwarschild
radius in the case of non-rotating BHs. Particles and photons can fall in through
this radius, but they cannot emerge outwards according to the
classical theory of GR. If the system has too much angular momentum J, no BH will be formed. The maximum angular momentum corresponds to $J=GM^2/c$. For a maximally rotating BH, the horizon radius is $r_{\rm +}=GM/c^2$, just half the result for the case of Schwarschild BH, $r_{\rm Sch} = 2GM / c^2$. For maximally rotating BHs the last stable circular orbit is
located at (Shapiro \& Teukolsky 1983) $r=r_{\rm +}=GM/c^2$ for corotating test particles and at $r=9r_{\rm +}=9GM/c^2$ for counter-rotating particles. (Recall: for non-rotating BHs, the last stable orbit is at $r=3r_{\rm Sch}=6GM/c^2$). Correspondingly, the maximum binding energies of these orbits can be found,
that is, the amount of energy that has to be lost in order for the material
to attain a bound stable circular orbit with radius r. In the corotating case, 42.3\% of the rest mass energy of the material can be released
as it spirals into the BH through a sequence of almost circular
equatorial orbits. In the counter-rotating case it is 3.8\%. This is the process by which energy is liberated through the accretion of matter
onto BHs, and is likely to be the source of energy in some of the most extreme
astrophysical objects. The energy available is much greater than that attainable from nuclear fusion
processes, which at most can release about 1\% of the rest mass energy of matter. A more complete treatment of the topics developed in this subsection can be found in Longair (1994a, 1994b).

\section{X-ray binaries}\label{xrb}

An X-ray binary is a binary system containing a compact object, either a
neutron star or a stellar mass black hole, that emits X-rays as a result of a process of accretion of matter from the companion star. Several scenarios 
have been proposed to explain this X-ray emission, depending on the nature of the compact object, its magnetic field in the case of a neutron star, and the geometry of the accretion flow. The accreted matter is accelerated to
relativistic speeds, transforming its potential energy provided by the intense gravitational field of the compact object into kinetic energy. Assuming that this kinetic energy is finally radiated, the accretion luminosity can be
computed, finding that this mechanism provides a very efficient source of
energy, even much higher efficiency than that for nuclear reactions.

On its way to the compact object, the accreted matter carries angular momentum
and usually forms an accretion disk around it. The matter in the disc looses
angular momentum due to viscous dissipation, which produces a heating of the
disc, and falls towards the compact object in a spiral trajectory. The black
body temperature of the last stable orbit in the case of a BH accreting at the
Eddington limit is given by:
\begin{equation}
T \sim 2 \times 10^7 M^{-1/4}
\label{eq:tlast}
\end{equation}
where $T$ is expressed in Kelvin and $M$ in M$_{\odot}$ (Rees 1984). For a
compact object of a few solar masses, $T \sim10^7$~K. At this temperature the energy is mainly radiated in the X-ray domain.

In High Mass X-ray Binaries (HMXBs) the donor star is an O or B early type
star of mass in the range $\sim8$--$20$~M$_{\odot}$ and typical orbital
periods of several days. HMXBs are conventionally divided into two subgroups:
systems containing a B star with emission lines (Be stars), and systems
containing a supergiant (SG) O or B star. In the first case, the Be stars do
not fill their Roche lobe, and accretion onto the compact object is produced
via mass transfer through a decretion disc. Most of these systems are
transient X-ray sources during periastron passage. In the second case, OB SG
stars, the mass transfer is due to a strong stellar wind and/or to Roche lobe
overflow. The X-ray emission is persistent, and large variability is common.
The most recent catalogue of HMXBs was compiled by Liu et al. (2006),
and contains 242 sources.

In Low Mass X-ray Binaries (LMXBs) the donor has a spectral type later than B,
and a mass $\leq2$~M$_{\odot}$. Although it is typically a non-degenerated star,
there are some examples where the donor is a WD. The orbital periods are in
the range 0.2--400 hours, with typical values $<24$ hours. The orbits are
usually circular, and mass transfer is due to Roche lobe overflow. Most
LMXBs are transients, probably as a result of an instability in the accretion
disc or a mass ejection episode from the companion. The typical ratio between
X-ray to optical luminosity is in the range $L_{\rm X}/L_{\rm
opt}\simeq100$--$1000$, and the optical emission is dominated by X-ray heating of
the accretion disc and the companion star. Some LMXBs are classified as `Z'
and `Atoll' sources, according to the pattern traced out in the X-ray
color-color diagram. `Z' sources are thought to be weak magnetic field neutron
stars of the order of $10^{10}$~G with accretion rates around
0.5--1.0~$\dot{M}_{\rm Edd}$. `Atoll' sources are believed to have even weaker
magnetic fields of $\leq10^8$~G and lower accretion rates of
0.01--0.1~$\dot{M}_{\rm Edd}$. The most recent catalogue of LMXBs was 
compiled by Liu et al. (2007), and contains 188 sources.

Recently, Grimm et al. (2002) estimated that the total number of
X-ray binaries in the Galaxy brighter than 2$\times 10^{34}$ erg~s$^{-1}$ is
about 705, being distributed as $\sim$325 LMXBs and $\sim$380 HMXBs.

\subsection{Radio emitting X-ray binaries (REXBs)}

The first X-ray binary known to display radio emission was Sco~X-1 in the late
1960s. Since then, many X-ray binaries have been detected at radio wavelengths
with flux densities $\geq0.1$--$1$~mJy. The flux densities detected are
produced in small angular scales, which rules out a thermal emission
mechanism. The most efficient known mechanism for production of intense radio
emission from astronomical sources is the synchrotron emission mechanism, in
which highly relativistic electrons interacting with magnetic fields produce
intense radio emission that tends to be linearly polarized. The observed radio
emission can be explained by assuming a spatial distribution of non-thermal
relativistic electrons, usually with a power-law energy distribution,
interacting with magnetic fields.

Since some REXBs, like SS~433, were found to display elongated or jet-like
features, as in Active Galactic Nuclei (AGN) and quasars, it was proposed that flows of relativistic
electrons were ejected perpendicular to the accretion disc, and were
responsible for synchrotron radio emission in the presence of a magnetic
field. Models of adiabatically expanding synchrotron radiation-emitting
conical jets may explain some of the characteristics of radio emission from
X-ray binaries (Hjellming \& Johnston 1988). Several models have
been proposed for the formation and collimation of the jets, including the
presence of an accretion disc close to the compact object, a magnetic field in
the accretion disc, or a high spin for the compact object. However, there is
no clear agreement on what mechanism is exactly at work.

There are eight radio emitting HMXBs and 35 radio emitting LMXBs. Since the
strong magnetic field of the X-ray pulsars disrupts the accretion disc at
several thousand kilometers from the neutron star, there is no inner accretion
disc to launch a jet and no synchrotron radio emission has ever been detected
in any of these sources. Although the division of X-ray binaries in HMXBs and
LMXBs is useful for the study of binary evolution, it is probably not
important for the study of the radio emission in these systems, where the only
important aspect seems to be the presence of an inner accretion disc capable
of producing radio jets. However, the eight radio emitting HMXBs include six
persistent and two transient sources, while among the 35 radio emitting LMXBs
we find 11 persistent and 24 transient sources. The difference between the
persistent and transient behavior clearly depends on the mass of the donor.

Excluding X-ray pulsars, $20$ to $25\%$ of the catalogued galactic X-ray
binaries have been detected at radio wavelengths regardless of the nature of
the donor. The corresponding ratio of detected/observed sources is probably
much higher. However, it is difficult to give reliable numbers, since
observational constrains arise when considering transient sources observed in
the past (large X-ray error boxes, single dish and/or poor sensitivity radio
observations, etc.), and likely many non-detections have not been published.

\subsection{Microquasars}

\begin{figure}  
\begin{center}
\hspace{0.25cm}
   \psfig{figure=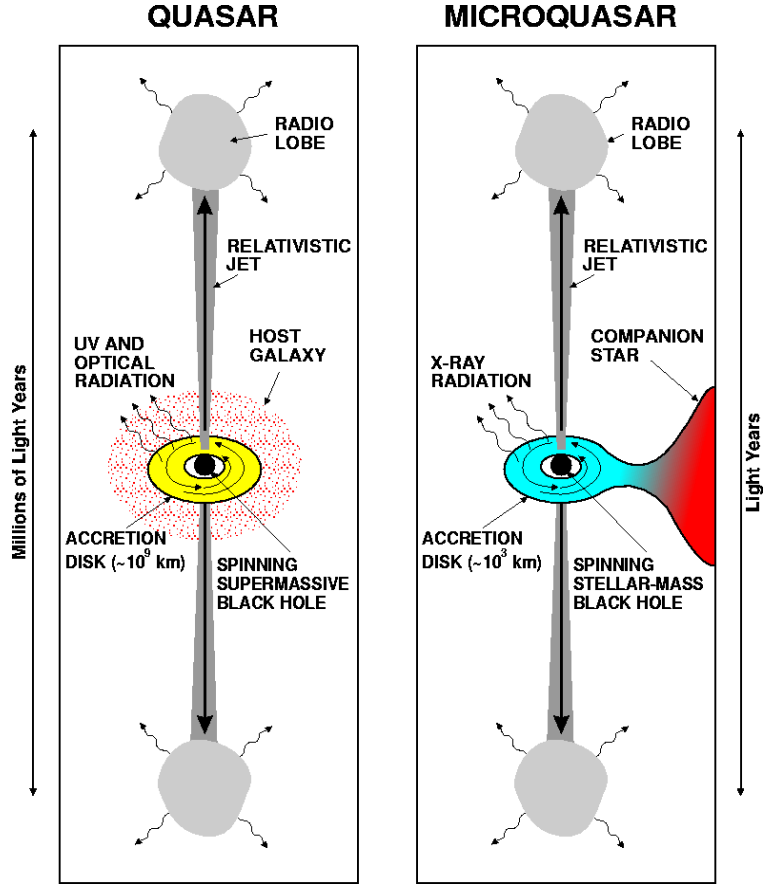,width=10.cm}
\caption{Comparative illustration of the analogy between quasars and microquasars. Note the extreme differences in the order of magnitude of the physical parameters involved. Figure reproduced from Mirabel \& Rodr\'{\i}guez (1998).}
\label{qmq}
\end{center}
\end{figure}

A microquasar is a radio emitting X-ray binary displaying relativistic radio
jets. The name was given not only because of the observed morphological
similarities between these sources and the distant quasars but also because of
physical similarities, since when the compact object is a black hole, some
parameters scale with the mass of the central object (Mirabel \&
Rodr\'{\i}guez 1999). A schematic illustration comparing some
parameters in quasars and microquasars is shown in Figure~\ref{qmq}.

From Eq.~\ref{eq:tlast}, a typical temperature of the accretion disc of a microquasar containing a
stellar mass black hole is $T\sim10^7$~K, while that of a quasar containing a
super-massive black hole ($10^7$--$10^9$~$M_{\odot}$) is $T\sim10^5$~K. This
explains why in microquasars the accretion luminosity is radiated in X-rays, 
while in quasars it is radiated in the optical/UV domain. The characteristic
jet sizes appear to be proportional to the mass of the black hole. Radio jets in
microquasars have typical sizes of a few light years, while in quasars may
reach distances of up to several million light years. The timescales are also
directly scaled with the mass of the black hole following $\tau\simeq R_{\rm
Sch}/c=2GM_{\rm X}/c^3\propto M_{\rm X}$. Therefore, phenomena that take place
in timescales of years in quasars can be studied in minutes in microquasars.
Thus, microquasars mimic, on smaller scales, many of the phenomena seen in
AGNs and quasars, but allow a better and faster progress in the understanding
of the accretion/ejection processes that take place near compact objects. 

The current number of confirmed microquasars is $\sim$16 among the 43 catalogued REXBs (Paredes 2005).
Some authors (Fender 2001) have proposed that all REXBs are microquasars, and
would be detected as such provided that there is enough sensitivity and/or
resolution in the radio observations. 

\subsubsection{Accretion disc and jet ejection}

The theoretical models attempting to understand the jet formation and its
connection with the accretion disc had a seminal contribution in the works by
Blandford \& Payne (1982). These authors explored the
possibility of extracting energy and angular momentum from the accretion disc
by means of a magnetic field whose lines extend towards large distances from
the disc surface. Their main result was the confirmation of the theoretical
possibility to generate a flow of matter outwards from the disc itself,
provided that the angle between the disc and the field lines was smaller than
60$^{\circ}$. Later on, the flow of matter is collimated at large distances
from the disc by the action of a toroidal component of the magnetic field. In
this way, two opposite jets could be formed flowing away perpendicularly to
the accretion disc plane.

\begin{figure}  
\begin{center}
\hspace{0.25cm}
   \psfig{figure=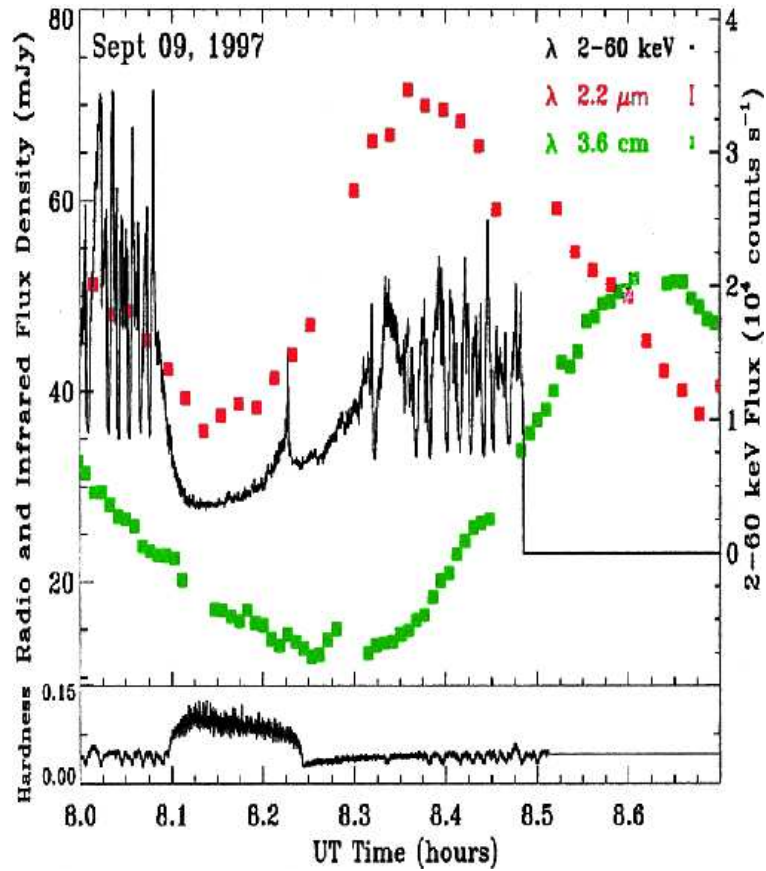,width=10.cm}
\caption{Multi-wavelength behaviour of the microquasar GRS~1915+105 as observed in September 8th 1997 (Mirabel et al. 1998, Chaty 1998). The radio data at 3.6~cm (grey squares) were obtained with the VLA interferometer; the infrared observations at 2.2 micron (black squares) are from the UKIRT; the continuous line is the X-ray emission as observed by RXTE in the 2--50 keV range. The hardness ratio (13--60 keV)/(2--13 keV) is shown at the bottom.}
\label{grs1915}
\end{center}
\end{figure}

To confirm observationally the link between accretion disc and the genesis of
the jets is by no means an easy task. The collimated ejections in GRS~1915+105
provide one of the best studied cases supporting the proposed disc/jet
connection. In Figure~\ref{grs1915}, from Mirabel et al. (1998),
simultaneous observations are presented at radio, infrared and X-ray
wavelengths. The data show the development of a radio outburst, with a peak
flux density of about 50~mJy, as a result of a bipolar ejection of plasma
clouds. However, preceding the radio outburst there was a clear precursor
outburst in the infrared. The simplest interpretation is that both flaring
episodes, radio and infrared, were due to synchrotron radiation generated by
the same relativistic electrons of the ejected plasma. The adiabatic expansion
of plasma clouds in the jets causes an energy loss in the electron population and,
as a result, the spectral maximum of their synchrotron radiation is
progressively shifted from the infrared to the radio domain.

It is also important to note the behaviour of the X-ray emission in
Figure~\ref{grs1915}. The emergence of jet plasma clouds, that produce the
infrared and radio flares, seems to be accompanied by a sharp decay and
hardening of the X-ray emission (8.08--8.23~h UT in the figure). The X-ray
fading is interpreted as the disappearance, or emptying, of the inner regions
of the accretion disc (Belloni et al. 1997). Part of the matter
content in the disc is then ejected into the jets, perpendicularly to the
disc, while the rest is finally captured by the central black hole.
Additionally, Mirabel et al. (1998) suggest that the initial time
of the ejection coincides with the isolated X-ray spike just when the hardness
ratio suddenly declines (8.23~h UT). The recovery of the X-ray emission level
at this point is interpreted as the progressive refilling of the inner
accretion disc with a new supply of matter until reaching the last stable
orbit around the black hole.

This behaviour in the light curves of GRS~1915+105 has been repeatedly
observed by different authors (e.g. Fender et al. 1997; Eikenberry et al. 1998), providing thus a solid proof of the so-called
disc/jet symbiosis in accretion discs. All the observed events were shorter than 1/2 h, and their equivalent in quasars, or AGNs, would require
a much longer minimum time span of some few years. Despite the complexity
in the GRS~1915+105 light curves, the episodes of X-ray emission decay with
associated hardening are reminiscent of the well known low/hard state typical
of persistent black hole candidates (Cygnus~X-1, 1E~1740.7$-$2942,
GRS~1758$-$258 and GX~339$-$4). The transitions towards this state are often
accompanied by radio emission with flat spectrum, interpreted as due to the
continuous creation of partially self-absorbed compact synchrotron jets.

It is worth mentioning the observational work by Marscher et al. (2002), who presented evidence of the disc/jet symbiosis also in the active galaxy 3C~120. Using VLBI techniques they observed episodes
of ejection of superluminal plasma just after the decay and hardening of the
X-ray emission. This is precisely the same behaviour displayed by
GRS~1915+105. The events in 3C~120 seem to be recurrent with an interval of
one year, which is consistent with a mass of the compact object of
$\sim10^{7}$ \mo. Such observations strongly support the idea of
continuity between galactic microquasars and AGNs in the Universe.

\section{The first BH candidates}

\begin{figure}  
\begin{center}
\hspace{0.25cm}
   \psfig{figure=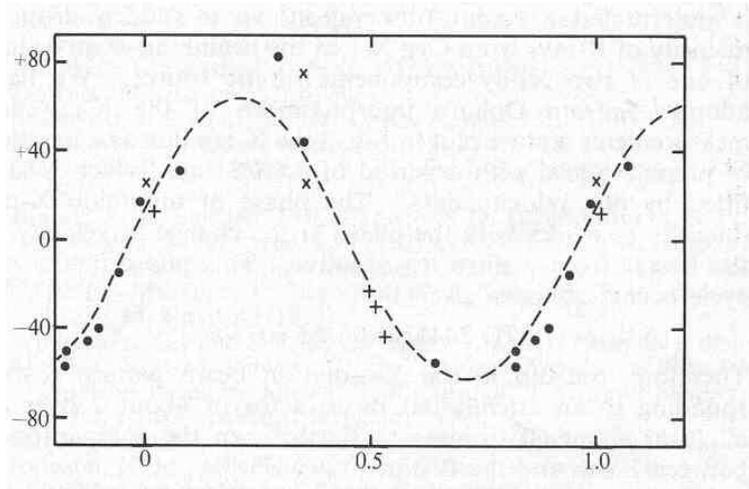,width=10.cm}
\caption{Radial velocity curve of 
HD 226868, the O9.7Iab companion star in the HMXB Cyg X-1, 
folded on the 5.6 day orbital period. Figure reproduced from Webster \& 
Murdin (1972).}
\label{cygx1}
\end{center}
\end{figure}

One of the first optical counterparts to be identified was the 9th magnitude supergiant star HD 226868, associated with the HMXB Cyg X-1. It showed radial velocity variations which made it a prime candidate for a stellar mass BH (Webster \& Murdin 1972, Bolton 1972). The supergiant star was shown to move with a velocity amplitude of ~64 \kms (later refined to 75 \kms) in a 5.6 day orbit due to the gravitational influence of an unseen companion (see Fig.~\ref{cygx1}). The mass function of the compact object was $f(M_{1}$) = 0.25 \,\mo. The mass of the bright companion $M_{2}$ is large and has a wide range of uncertainty. 
If the optical star were a normal O9.7Iab its mass would be ~33 \mo\, which, for an edge-on orbit ($i$ = 90\gr) would imply a compact object of $\sim$ 7 \mo.
However, the optical star is likely to be undermassive 
for its spectral type as a result of mass transfer and binary evolution, as has been shown to be the case in several NS binaries (e.g. Rappaport \& Joss 1983). 

It could be undermassive by as much as a factor of 3 given the uncertainty in distance, log $g$ and $T_{\rm eff}$. A plausible lower limit of 10 \mo, combined with an upper limit to the inclination of 60\gr, based on the absence of X-ray eclipses, leads to a compact object of $>$ 4 \mo (Bolton 1975).

In 1975, the X-ray satellite Ariel V detected A0620-00. This source 
is a X-ray transient (XRT), a subclass of LMXBs which undergo dramatic episodes of enhanced mass-transfer or "outbursts" triggered by viscous-thermal instabilities in the disc.
During outburst, the companion remains undetected because it is totally overwhelmed by the intense optical light from the X-ray heated disc. 
After a few months of activity the X-rays switch off, the reprocessed flux drops several magnitudes into quiescence and the companion star becomes the dominant source of optical light. This offers the opportunity to perform radial velocity studies of the cool companion and unveil the nature of the compact star. 

The first detection of the companion in A0620-00 revealed a mid-K star moving in a 7.8 hr period with velocity amplitude of 457 \kms. 
The implied mass function was 3.2 $\pm$ 0.2 \mo, the largest ever measured (McClintock \& Remillard 1986). A lower limit to the mass of the compact star of 3.2 \mo was established by assuming a very conservative low-mass companion of 0.25 \mo and $i$ $<$ 85\gr, based on the lack of X-ray eclipses. This exceeds the maximum mass allowed for a stable NS and hence it also became a very compelling case for a BH.

In the 1980s there was a hot debate about the real existence of BHs. 
There were three firm candidates, two HMXBs (Cyg X-1 and LMC X-3) and a transient LMXB (A0620-00),  all with lower limits to $M_{1}$ very close to the maximum mass for NS stability. Alternative scenarios were proposed to avoid the need for BHs such as multiple star systems (Bahcall et al. 1974) or non-standard models invoking exotic equations of state (EoS) for condensed matter. It was proposed that the "Holy Grail in the search for BHs is a system with a mass function that is plainly 5 \mo\ or greater" (McClintock 1986).

\begin{figure}  
\begin{center}
\hspace{0.25cm}
   \psfig{figure=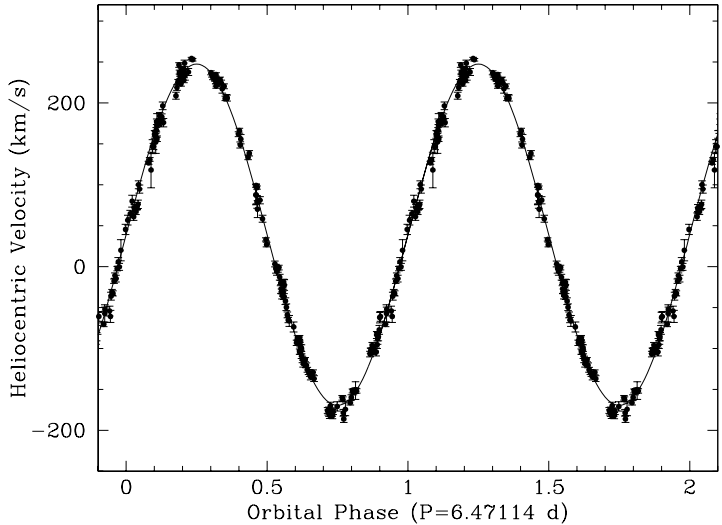,width=10.cm,angle=-90}
\caption{Radial velocity curve of the K0 companion in the transient LMXB 
V404 Cyg during quiescence. This graph contains velocity points obtained
between 1991 and 2005 (Casares et al. 1992, 2007).}
\label{v404}
\end{center}
\end{figure}

In 1989, Ginga discovered a new XRT in outburst named GS2023+338 (= V404 Cyg). Its X-ray properties drew considerable attention because of the exhibition of a possible luminosity saturation at L$_{\rm X}$ = $10^{39}$ erg s$^{-1}$ and dramatic variability (Zycki et al. 1999). Spectroscopic analysis during quiescence revealed a K0 star moving with a velocity amplitude of 211\,km s$^{-1}$ in a 6.5 day orbit (see Fig.~\ref{v404}).
The obtained mass function, $f(M_{1}$) = 6.08 $\pm$ 0.06 \mo, implies the 
compact object must be more massive than 6 \mo (Casares et al. 1992).

\subsection{BH binaries in the Milky Way}

The mass function of V404 Cyg was the highest yet measured and, very clearly, placed an accreting compact object comfortably above the upper limit of maximally rotating NSs for any "standard" EoS assumed.
This work on V404 Cygni revolutionized the field of BH searches and this source is widely considered as the best evidence for a BH, where no additional assumptions on $i$ nor $M_2$ have to be invoked.

This remarkable result obtained by Casares and collaborators established V404 Cyg as the "Holy Grail" BH for almost a decade. 

Since then, many other BHs have been unveiled through dynamical studies of XRTs in quiescence, some of them having mass functions also in excess of 5\,\mo. This has been possible thanks to the improvement of the instrumentation and the new generation of 10-m class telescopes. The list of the confirmed stellar BHs has increased recently with the addition of two stellar BHs. One of them, M33 X-7, is a  15.65 $\pm$ 1.45 \mo\, BH in an eclipsing binary in the nearby spiral galaxy M33 (Orosz et al. 2007). The other is IC~10 X-1, located in the Local Group starburst galaxy IC~10, and with a mass of 32.7 $\pm$ 2.6 \mo, being indeed the most massive known stellar mass BH (Silverman \& Filippenko 2008). In Table~\ref{tabbh} is presented a compilation of the all confirmed stellar BHs made by Casares (2007), with the addition of these two new BHs.

\begin{table}\def~{\hphantom{0}}
  \begin{center}
  \caption{Confirmed black holes and mass determinations. Table from Casares (2007) updated with M33 X-7 and IC 10 X-1.} 
  \label{tabbh}
  \begin{tabular}{lccccc}\hline
 System &  $P_{\rm{}orb}$ &  $f(M)$ & Donor  &  Classification & $M_{\rm x}$ \\
 &   [days] &  [M$_{\odot}$] &  Spect. Type & &  [M$_{\odot}$] \\ \hline
GRS 1915+105    &     33.5    &    9.5 $\pm$ 3.0        &    K/M III   & LMXB/Transient   &   14 $\pm$ 4   \\
V404 Cyg        &      6.471  &   6.09 $\pm$ 0.04       &    K0 IV     &      ,,          &   12 $\pm$ 2   \\
Cyg X-1         &      5.600  &  0.244 $\pm$ 0.005      &    09.7 Iab  &  HMXB/Persistent &   10 $\pm$ 3   \\
M33 X-7$^a$     & 3.453       &          --              &    O7 III     &    --
                & 15.65 $\pm$ 1.45  \\
LMC X-1         &      4.229  &   0.14 $\pm$ 0.05       &    07 III    &      ,,          &   $>$ 4         \\
XTE J1819-254   &      2.816  &   3.13 $\pm$ 0.13       &    B9 III    &  IMXB/Transient  &  7.1 $\pm$ 0.3 \\ 
GRO J1655-40    &      2.620  &   2.73 $\pm$ 0.09       &    F3/5 IV   &      ,,          &  6.3 $\pm$ 0.3 \\
BW Cir &    2.545  &   5.74 $\pm$ 0.29       &    G5 IV     &  LMXB/Transient  &    $>$ 7.8     \\	 
GX 339-4        &      1.754  &   5.8  $\pm$ 0.5        &     --       &      ,,          &       --         \\
LMC X-3         &      1.704  &   2.3  $\pm$ 0.3        &    B3 V      &  HMXB/Persistent &  7.6 $\pm$ 1.3 \\
XTE J1550-564   &      1.542  &   6.86 $\pm$ 0.71       &    G8/K8 IV  &  LMXB/Transient  &  9.6 $\pm$ 1.2 \\
IC~10 X-1$^b$   &    1.455   &   7.64 $\pm$ 1.26           &   --    &   Wolf-Rayet     &  32.7 $\pm$ 2.6  \\
4U 1543-475     &      1.125  &   0.25 $\pm$ 0.01       &    A2 V      &  IMXB/Transient  &  9.4 $\pm$ 1.0 \\
H1705-250       &      0.520  &   4.86 $\pm$ 0.13       &    K3/7 V    &  LMXB/Transient  &    6 $\pm$ 2   \\
GS 1124-684     &      0.433  &   3.01 $\pm$ 0.15       &    K3/5 V    &      ,,          &  7.0 $\pm$ 0.6 \\
XTE J1859+226   &  0.382  &   7.4  $\pm$ 1.1        &     --       &      ,,          &          --      \\
GS2000+250      &      0.345  &   5.01 $\pm$ 0.12       &    K3/7 V    &      ,,          &  7.5 $\pm$ 0.3 \\
A0620-003       &      0.325  &   2.72 $\pm$ 0.06       &    K4 V      &      ,,          &   11 $\pm$ 2   \\
XTE J1650-500   &      0.321  &   2.73 $\pm$ 0.56       &    K4 V      &      ,,          &      --          \\
GRS 1009-45     &      0.283  &   3.17 $\pm$ 0.12       &    K7/M0 V   &      ,,          &  5.2 $\pm$ 0.6 \\
GRO J0422+32    &      0.212  &   1.19 $\pm$ 0.02       &    M2 V      &      ,,          &    4 $\pm$ 1   \\
XTE J1118+480   &      0.171  &   6.3  $\pm$ 0.2        &    K5/M0 V   &      ,,          &  6.8 $\pm$ 0.4 \\\hline
  \end{tabular}
  \end{center}
$^a$ Orosz et al. (2007). \\
$^b$ Silverman \& Filippenko (2008).

\end{table}
%

\begin{figure}  
\begin{center}
\hspace{0.25cm}
   \psfig{figure=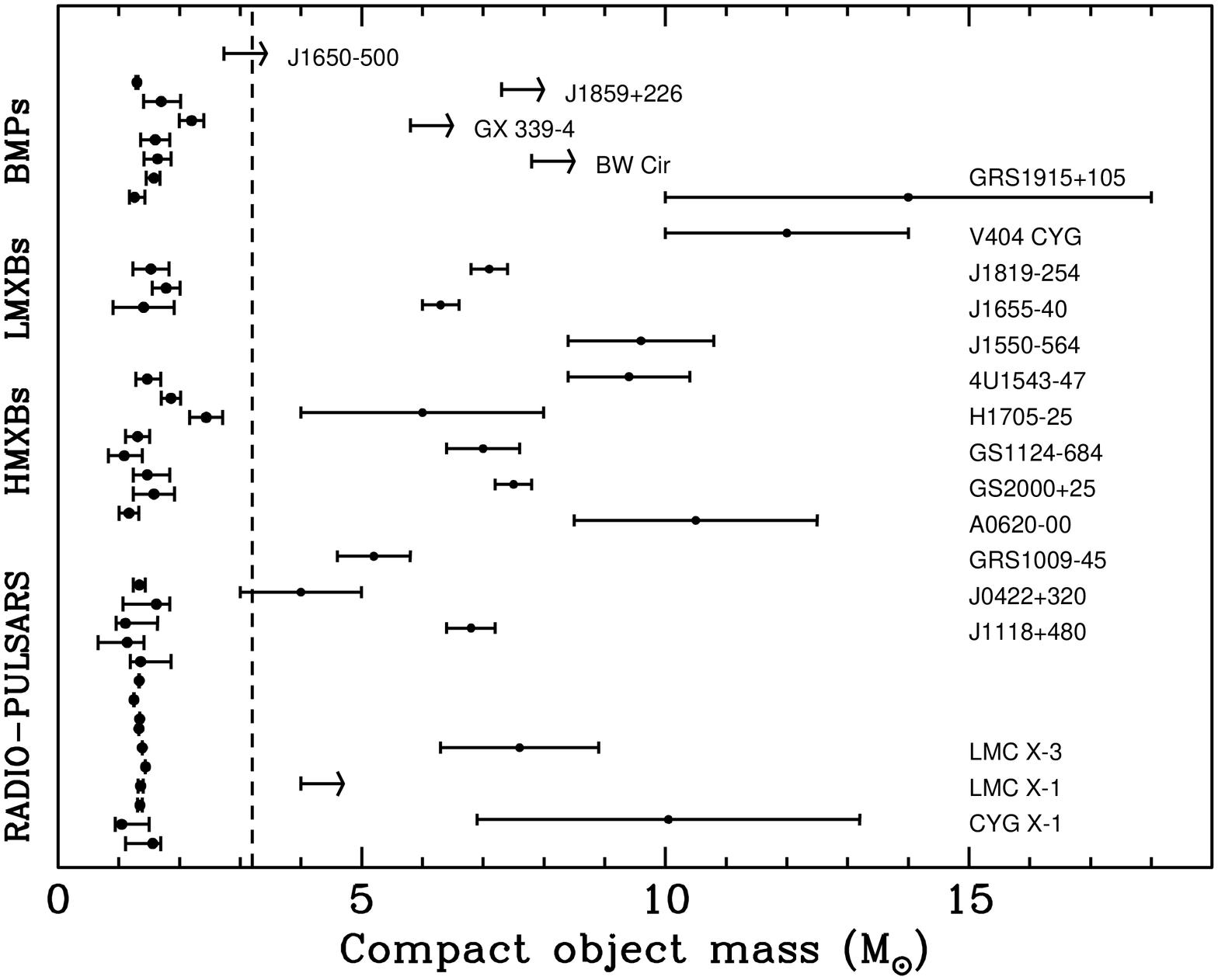,width=10.cm,angle=-90}
\caption{Mass distribution of compact objects in X-ray binaries. 
  Arrows indicate lower limits to BH masses. Figure reproduced from Casares (2007).}
\label{distribution}
\end{center}
\end{figure}

\section{Observational signatures of the BHs}

\subsection{Mass determination}
A complete determination of the component masses in a binary system
requires the radial velocity curves of both stars and a knowledge of the
inclination angle.
The determination of masses in LMXBs is usually reduced to a single line
spectroscopic binary problem where all the information has to be
extracted from the optical star.
A complete solution to the system parameters can be obtained from
three observational experiments involving high resolution optical
spectroscopy and infrared photometry (the mass function, rotational broadening and ellipsoidal modulation).

\subsubsection{The mass function.}

This is a fundamental equation for the determination of binary system
parameters which can be derived directly from Kepler's 2nd and 3rd laws. It
relates the masses of the two stars $(M_1, M_2)$ and the inclination angle ($i$)
through two observable quantities which are readily measured from the
radial velocity curve of the optical star: the orbital period $P_{\rm orb}$ and the
radial velocity semi-amplitude $K_2$. 

\begin{equation}
f(M_1)=\frac{K_{2}^{3}\,P_{\rm orb}}{2\pi G}=\frac{M_{1}^{3}\,{\rm sin}^3\,i}{(M_1+M_2)^2}=\frac{M_1\,{\rm sin}^3\,i}{(1+q)^2}
\end{equation}

where $q=M_2/M_1$ is the mass ratio.
$f(M_1)$ represents an absolute lower limit to $M_1$,
i.e.~at the extreme case when the mass of the
companion star is neglected (i.e.~$q \rightarrow 0$) and
the binary is seen edge on ($i = 90^{\circ})$. However, $f(M)\propto K_{2}^{3}$ implies thar the mass function is extremely sensitive to uncertainties in the radial velocity amplitude. Any non-uniform brightness distribution across the surface of the companion star
will modify the radial velocity curve and affect our system parameter determination.

A source of systematic error in the determination of $K_2$ is produced by the X-ray irradiation, which tends to suppress the absorption lines near the L1 point. The consequence of the X-ray irradiation is the displacement of the centre of light away from the companion's centre of mass and the radial velocity curve will be significantly eccentric. A sine-wave fit to it will
give an observed $K_{\rm obs}$ which will be larger than the true $K_2$. Irradiation is considered to be negligible in quiescent X-ray transients
(where $L_x \leq 10^{33}$ erg s$^{-1})$ but it can be important in X-ray active states. An example of the importance of this effect was given by GRO J1655$-$40 (=N Sco 94). During the decay of the 1994 outburst ($L_x = 1.4\times 10^{37}$ erg s$^{-1}$) the radial velocity
curve was fitted with a simple sine-wave, obtaining $K_2 = 228.2 \pm 2.2$ km s$^{-1 }$. This, combined with $P = 2.6$ d gave $f(M) = 3.24 \pm$ 0.09M$_{\odot}$ and hence was a strong case for a BH (Orosz \& Bailyn 1997). The same data was subsequently fitted by other authors using an irradiation model,
obtaining $K_2 =$ 192 -- 214 km s$^{-1}$ which reduces the mass function to
$f(M) =$ 1.93 -- 2.67 M$_{\odot}$ (Phillips et al. 1999). This latter result was finally confirmed by observations in true quiescence which
give $K_2 = 215.5 \pm 2.4$ km s$^{-1}$ and $f(M) = 2.73 \pm$ 0.09M$_{\odot}$  (Shahbaz et al. 1999). The new mass function is 16\% lower than the first reported value, enough to
disclaim GRO J1655-40 as a secure BH candidate.

\subsubsection{Rotational broadening.}

The companion stars transfer matter onto their compact objects and hence
they must be filling their Roche lobes. In addition, the short orbital periods
($\sim$ hr) and old ages (> 10$^7$ yr) of these close binaries suggest that the
companion stars must be synchronised, i.e. $\omega_{\rm s} = \omega_{\rm orb}$ where $\omega_{\rm s}$ and $\omega_{\rm orb}$
are the stellar and orbital angular velocities respectively. The projected
linear velocities would then be

\begin{equation}
\frac{V_{\rm rot}\,{\rm sin}\,i}{R_2}=\frac{K_2+K_1}{a}=\frac{K_2\,(1+q)}{a}
\end{equation}

where $q=M_2/M_1=K_1/K_2$, $R_2$ is the equivalent radius of the companion's Roche lobe and 
$a$ is the binary separation.
Combining with $\frac{R_2}{a}\simeq0.462\Big(\frac{q}{1+q}\Big)^{1/3}$ one obtains $V_{\rm rot}\,{\rm sin}\,i=0.46 K_2\,q^{1/3}\,(1+q)^{2/3}$. Therefore, the mass ratio $q$ can be measured directly from the radial velocity curve
and the observed rotational broadening ($V_{\rm rot}\,{\rm sin}\,i$) of the secondary's
absorption lines (Gies \& Bolton 1986). The rotational velocity is determined by comparing our target with
broadened versions of spectral type templates (e.g. through a $\chi^2$
minimization technique) (Marsh et al. 1994). An example of rotational broadening analysis, applied to V404 Cygni, can be found in Casares \& Charles (1994).	

\subsubsection{Ellipsoidal modulation.}

The lightcurves of companion stars in contact binaries display the characteristic
double-humped variation on the orbital period, with amplitudes $\le10\%$.
This modulation is a consequence of the tidal distortion of the secondary star
and the non-uniform distribution of the surface brightness, due to a combination
of limb and gravity darkening. The dependence of $T_{\rm eff}$ on the local surface gravity $g$ implies that the inner
hemisphere of the secondary (facing the compact object) will be cooler and,
therefore, the phase 0.5 minimum becomes deeper than the minimum at
phase 0. Model calculations show that the shape and amplitude of the ellipsoidal modulation are functions of $q$ and $i$.
In particular, the amplitude is a strong (increasing) function of $i$,
and is insensitive to $q$ if $q \le 0.1$. This is normally the case in SXTs and therefore model fits to the ellipsoidal modulation can be used to determine $i$ directly.

\subsubsection{Fluorescence emission from the irradiated donor}

There are 20 XRBs that lack radial velocity data. Most of them even lack an
optical counterpart, and only three have known orbital periods. Nevertheless,
they are considered black-hole candidates because they closely resemble black hole binaries
in their X-ray spectral and temporal behavior (Remillard \& McClintock 2006).
Unfortunately, they have never been seen in quiescence, or they simply become
too faint for an optical detection of the companion star. Fortunately, a 
new technique to extract dynamical information during their X-ray
active states has been applied recently. It utilises narrow high-excitation emission lines powered by irradiation on the companion star, in particular the strong CIII and fluorescence NIII lines from the Bowen blend at $\lambda\lambda$4630--40. This technique was first applied to the NS LMXB Sco X-1 and the Doppler shift
of the CIII/NIII lines enabled the motion of the donor star to be traced for the first
time (Steeghs \& Casares 2002). This was also attempted during the
2002 outburst of the BH candidate GX 339-4, using high-resolution
spectroscopy to resolve the sharp NIII/CIII lines. The orbital solution yields a velocity semi-amplitude of 317 km s$^{-1}$ which defines a strict lower limit to the velocity amplitude of the companion star because these lines arise from the irradiated hemisphere and not the centre of mass of the donor. A solid lower bound to the mass function is 5.8 M$_{\odot}$ which provides
compelling evidence for a BH in GX 339$-$4 (Hynes et al. 2003). This technique opens an avenue to extract dynamical information
from new XRTs in outburst and X-ray persistent LMXBs, which
hopefully will help increase the number of BH discoveries.

\subsection{Event horizon}

In order to show that a dynamical BH candidate (i.e., a massive compact object)
is a genuine BH, one would hope to demonstrate that the candidate has an
event horizon - the defining characteristic of a BH. The first test for the presence of the event horizon was carried out by Narayan et al. (1997) using archival X-ray data. One had to wait for Chandra before one could claim definitive evidence. Strong evidence has been
obtained for the reality of the event horizon from observations that compare
BHBs with very similar NS binaries. NS binaries show signatures of the hard
surface of a NS that are absent for the BH systems. \\

\subsubsection{BH are fainter than NS binaries (quiescence)}

\begin{figure}  
\begin{center}
\hspace{0.25cm}
   \psfig{figure=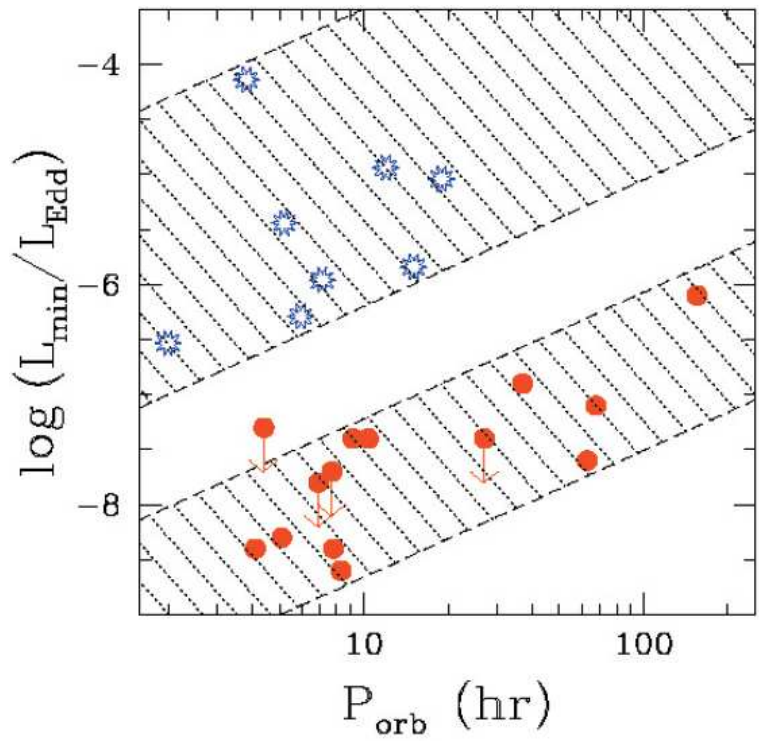,width=10.cm}
\caption{BH (red circles) and NS (blue stars) X-ray transients in
quiescence. Figure reproduced from McClintock et al. (2007)}
\label{chandra}
\end{center}
\end{figure}

BH systems are about 100 times fainter than the nearly identical NS
binaries. In Figure~\ref{chandra}, there are plotted the Eddington-scaled luminosities of BH (circles) and NS (stars) X-ray transients in quiescence (McClintock et al. 2007). At every orbital period, BH candidates have Eddington-scaled luminosities 
lower than NS by two to three orders of magnitude. The orbital period is a predictor of $\dot{M}$, according to binary mass transfer models (Menou et al. 1999). By comparing quiescent BHs and NSs with similar orbital periods, the uncertainty in the mass accretion rate is eliminated (Garcia et al. 2001). The explanation for this result is that accretion in these quiescent systems
occurs via a radiatively inefficient mode (advection-dominated accretion), as
confirmed through spectral studies (e.g, Narayan et al. 2002). Therefore,
the disc luminosity $L_{\rm disc} \ll \dot{M}c^2$ and most of the binding energy that is released as gas falls into the potential well is retained in the gas as thermal energy. In the case of an accreting NS,
this energy is eventually radiated from the stellar surface, and so an external
observer still sees the full accretion luminosity $\sim0.2 \dot{M}c^2$. In the case of a BH, however, the superheated gas falls through the event horizon, carrying all its thermal energy with it. The observer therefore sees only the disc luminosity $L_{\rm disc,}$ which is extremely small.

\subsubsection{BHs lack a soft thermal component (quiescence)}

BHs lack a soft thermal component of emission that is very
prevalent in the spectra of NSs and can be ascribed to surface
emission (McClintock et al. 2004).

\subsubsection{Type I thermonuclear bursts and lack of pulses}

This has been quantified using observations of dynamical BHs over 9
years of RXTE data. The probability that the non-detection of bursts were
consistent with a solid surface is found to be $\sim2\times10^{-7}$ (Remillard et al. 2006)

\subsubsection{High-frequency timing noise} 

The study of the Fourier power spectra at high frequencies offers a way to distinguish a NS from a BH (Sunyaev \& Revnivtsev 2000). After the
analysis of the power density spectra (PDS) of a sample of 9 NS and 9 BH binaries in the
low/hard spectral state, it is concluded that in the accreting NS with a weak magnetic field, significant power is contained at
frequencies close to one kHz. In the case of the accreting BH, the 
PDS show strong decline at frequencies higher than 10 -- 50 Hz. These empirical phenomenology could be used to help to distinguish the accreting NS from BH in X-ray transients.
The XRT with noise in their X-ray flux at frequencies higher than 500 Hz should be considered NS. Sunyaev \& Revnivtsev 
propose to explain the observed difference as a result of the existence of a
radiation dominated spreading layer on the NS surface.

\subsubsection{A distinctive spectral component from a boundary layer} 

The classic colour-colour diagram of XRBs at high accretion rates
shows a clear separation in the evolution of NS and BH binaries. This has
been ascribed to the presence of a boundary layer in NS which gives rise to
an additional thermal component in the spectrum and drags NS outside the
BH region (Done \& Gierlinski 2003).\\

All approaches to this subject can provide only indirect evidence
for the event horizon because it is quite impossible to detect directly
any radiation from this immaterial surface of infinite redshift.
Nevertheless, barring appeals to very exotic physics, the body of
evidence just considered makes a strong case that dynamical BH
candidates possess an event horizon.

\subsection{Measuring Black Hole Spin}

An astrophysical BH is described by two parameters, its mass $M$ and its
dimensionless spin parameter $a_*$. Because the masses of 21 BHs have
already been measured or constrained, the next obvious goal is to
measure spin. Several methods have been used to measure the BH spin. The first method, based on X-ray polarimetry, appears very promising but thus far has not
been incorporated into any contemporary X-ray mission. The second method, based on X-ray continuum fitting, is already producing useful results. The third method, based on the Fe K line profile, has also yielded results, although the
method is hampered by significant uncertainties. The fourth method, based on high-frequency QPOs, offer the most reliable measurement of spin once a model is established. Here follows some details of these methods. 

\subsubsection{X-ray polarimetry.}
The polarization features of BH disc radiation can be affected strongly by GR effects. Conventional disc models predict that higher energy photons come from smaller disc radii. Then as the photon energy increases
from 1 keV to 30 keV, the plane of linear polarization swings smoothly through an angle of about 40$^{\circ}$ for a 9M$_{\odot}$ Schwarzschild BH and 70$^{\circ}$ for an extreme Kerr BH (Connors et al. 1980). The effect is due to the strong gravitational bending of light rays.
In the Newtonian approximation, the polarization angle does not vary with energy. A gradual change of the plane of polarization with energy is a
purely relativistic effect, and the magnitude of the change can give a direct
measure of $a_*$. Polarimetry data do not even require $M$, although knowledge of $i$ is useful in order to avoid having to include that parameter in the fit.

\subsubsection{X-ray continuum fitting.}

This technique, determines the radius $R_{\rm in}$ of the inner edge of the accretion disc and assumes that this radius corresponds to $R_{\rm ISCO}$ (innermost stable circular orbit). Because $R_{\rm ISCO}/R_{\rm g}$ is a monotonic function of $a_*$, a measurement of  $R_{\rm in}$ and $M$
directly gives $a_*$.
$R_{\rm in}$ can be estimated provided that: (a) $i$ and $D$ are sufficiently well known, (b) the X-ray flux and spectral temperature are measured from well-calibrated X-ray data in the thermal state and, (c) the disc radiates as a blackbody.
The continuum fitting is the best current method for measuring spin, although its application requires accurate estimates of BH mass ($M$), disc inclination ($i$), and distance.

\subsubsection{The Fe K line profile.}

The first broad Fe K$_{\alpha}$ line observed for either a BHB or an AGN was reported in the spectrum of Cyg X$-$1 based on EXOSAT data (Barr et al. 1985).
The Fe K fluorescence line is thought to be generated through the irradiation
of the cold (weakly-ionized) disc by a source of hard X-rays (likely an optically thin, Comptonizing corona). In the case that the disc rotates around a BH and extends down to the ISCO, i.e. only a few gravitational radii away from the event horizon, relativistic beaming and gravitational redshifts in the inner disc region can
serve to create an asymmetric line profile (for a review, see Reynolds \& Nowak 2003). In 1995, the first relativistically broadened iron K profile was  observed in the
Seyfert--1 galaxy MCG--6--30--15 with the Japanese X-ray observatory ASCA (Tanaka et al. 1995). 

\begin{figure}  
\begin{center}
\hspace{0.25cm}
   \psfig{figure=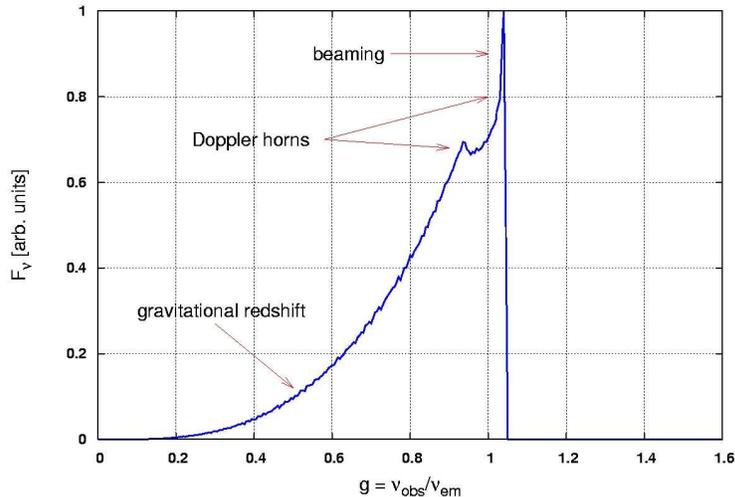,width=10.cm}
\caption{Example of a relativistic emission line profile that is emitted from a thin accretion disc around a Kerr BH with $a/M = 0.998$. The line profile is subject to the Doppler effect due to rotation of the disc (causes two
horns); special relativistic beaming (intensifies the blue line wing), and gravitational redshift (smears out
the whole line profile and produces an extended red line wing). Figure reproduced from Mueller (2007)}
\label{muller03}
\end{center}
\end{figure}

\begin{figure}  
\begin{center}
\hspace{0.25cm}
   \psfig{figure=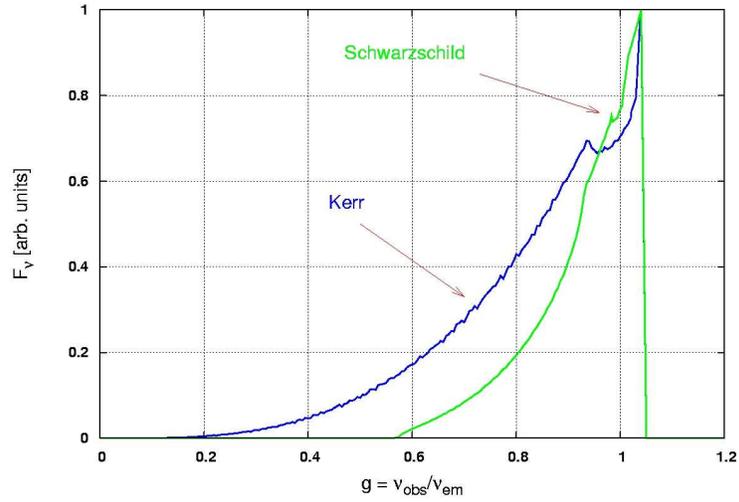,width=10.cm}
\caption{Comparison of observed relativistic
emission line profiles around a Kerr ($a/M = 0.998$, blue) and a Schwarzschild BH ($a/M
= 0$, green). Figure reproduced from Mueller (2007)}
\label{muller05}
\end{center}
\end{figure}

\begin{figure}  
\begin{center}
\hspace{0.25cm}
   \psfig{figure=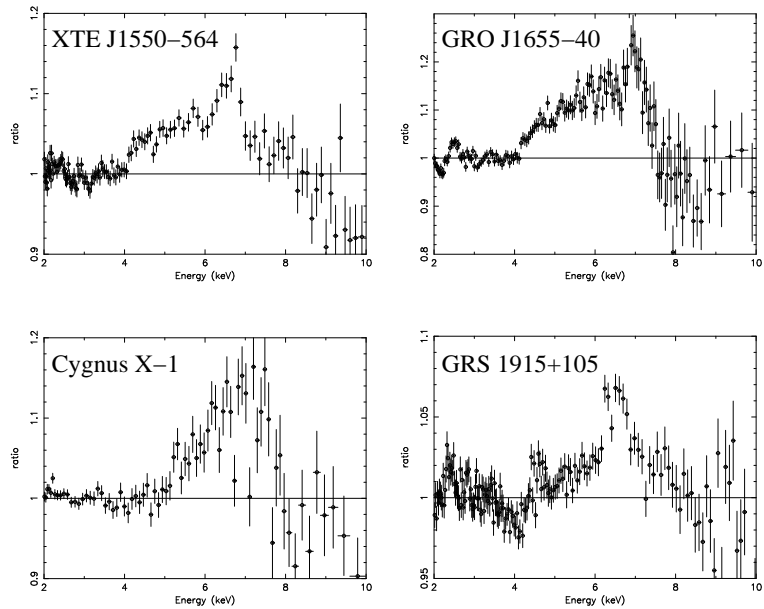,width=10.cm}
\caption{Relativistic X-ray lines from the inner accretion discs around four BHs. Figure reproduced from Miller (2007)}
\label{fourbh}
\end{center}
\end{figure}

The sharp line profile in the rest frame of the emitter (e.g.~an accretion disc) would be very different as observed in a distant laboratory frame. This is due to: 
1) The Doppler effect produces a redshift at the part that is receding
and a blueshift at the part of the disc that is approaching to the
observer. The result is a broadened line maybe with two Doppler peaks if
the disc is sufficiently inclined towards the observer (this already happens in
Newtonian physics).
2) For intermediate to large inclinations there is a special relativistic beaming effect at work. The Keplerian velocities are indeed comparable to the speed of light so that radiation is bent towards the observer. Therefore, the
observed radiation intensity is amplified as compared to the rest frame.
3) The BH captures line photons by strong space--time curvature. This
gravitational redshift effect is two--fold: the observed photon energy is
redshifted relative to the rest frame energy and the spectral line flux is lower in the observer's frame. All these three effects produce an asymmetric and skewed relativistic emission line profile (see Figure~\ref{muller03}). 
In addition, relativistic emission line profiles also indicate BH spin. It is known from accretion theory that standard discs extend down to the ISCO. The
ISCO depends on BH spin and is closer to the event horizon for rapidly spinning
Kerr BHs, $r_{\rm ISCO}=1r_{\rm g}$, as compared to Schwarzschild BHs, $r_{\rm ISCO}=6r_{\rm g}$. As a
consequence the line emission comes from regions that are deeper in the
gravitational potential if the BH rotates. Therefore, the line profiles from the vicinity
of Kerr BHs are more influenced by gravitational redshift (see Figure~\ref{muller05}). The line has been modeled in the spectra of several BHBs. In Figure~\ref{fourbh}, relativistic X-ray lines from the inner accretion discs around four BHs are shown (Miller 2007). GRS 1915+105 is clearly not as skewed as the others, and does not strongly require BH spin. In this source there has been some controversy about the value of the spin, being considered a rapidly rotating Kerr BH in one case, $a_*>0.98$ (McClintock et al. 2006) and an intermediate spin, $a_*\sim0.7$ in the other case (Middleton et al. 2006). 

In V4641 Sgr, the inner disc
radius deduced from the line profile is consistent with the $6R_{\rm g}$ radius of the
ISCO of a Schwarzschild BH, suggesting that rapid spin is not required (Miller et al. 2002). 

In GX 339$-$4, the inner disc likely extends inward to (2 -- 3)$R_{\rm g}$,
implying $a_*\ge 0.8 - 0.9$ (Miller et al. 2004). 
XTE J1650$-$500 is the most extreme case with the inner edge located at $\approx 2 R_{\rm g}$,
which suggests nearly maximal spin (Miller et al. 2002; Miniutti et al. 2004).
Large values of $a_*$ have also been reported for XTE J1655$-$40 and XTE
J1550$-$564 (Miller 2007).
Broadened Fe K lines data do not even require $M$ to obtain the spin parameters, although knowledge of $i$ is
useful in order to avoid having to include that parameter in the fit.

\subsubsection{High-frequency QPOs}

High-frequency quasi periodic oscillations (HFQPO) are likely to offer the most reliable
measurement of spin once the correct model is known. 
Typical frequencies of these fast QPOs, e.g., 150 -- 450 Hz, correspond 
respectively to the frequency at the ISCO for Schwarzschild BHs with masses of 
15 -- 5$\,M_{\odot}$, 
which in turn closely  matches the range of observed masses. These QPO 
frequencies (single or pairs) do not vary significantly despite sizable changes in the X-ray luminosity. This suggests that the frequencies are primarily 
dependent on the mass and spin of the BH. Those BHs that show HFQPOs and have 
well-constrained masses are the best prospects for constraining the value of the BH spin ($a_*$). Theoretical work aimed at explaining HFQPOs is motivated by three sources: GRO~J1655$-$40, XTE~J1550$-$564 and GRS~1915+105. 
These sources are presently the only ones that both exhibit harmonic (3:2) HFQPOs and have 
measured BH masses. If these HFQPOs are indeed GR oscillations,
then it would be possible that the three BHs have similar values of the spin
parameter $a_*$. It is relevant to try to confirm this 
result  by taking frequency and mass measurements for more sources.
Assuming we have a well-tested model, 
QPO observations only require knowledge of $M$ to provide a spin estimate.

\section{A super-massive Black Hole in the Milky Way}\label{gcb}

\subsection{Galaxies and BHs}

BH with masses of $10^6$ to few $10^9$\,M$_{\odot}$ are 
believed to be the engines that power nuclear activity in galaxies.
AGN range from faint, compact radio sources like that in M31 to quasars like 
3C 273 that are brighter than the whole galaxy in which they live. 
Some nuclei fire jets of energetic particles millions of lightyears into space. It is believed that this enormous flow of energy comes from the stars 
and gas that are falling into the central BH. 

In the past years, the search for super-massive BHs has been done by 
measuring rotation and random velocities of stars and gas near galactic centres. 
If the velocities are large enough,  then they imply more mass than we see 
in stars, being a BH the most probable explanation. More than 50 have been found, with masses in the range expected for nuclear engines, and consistent with predictions based on the energy output of quasars. 

The powerful gravitational force exerted by a super-massive BH in a galactic nucleus on nearby gas and stars, causes them to move at high speeds. This is difficult to see 
in quasars, because they are far away and because the dazzling light of the 
active nucleus swamps the light from the host galaxy. However, in radio galaxies with fainter nucleus, the stars and gas are more visible. 
The giant elliptical galaxy Messier 87, one of the two brightest objects in the 
Virgo cluster of galaxies, is a radio galaxy with a bright jet emerging from its 
nucleus. It has long been thought to contain a BH. Observations of Messier 87 
with the HST revealed a disc of gas 500 lightyears in diameter, whose orbital 
speeds imply a central mass of $3\times10^9$M$_{\odot}$. 
The ratio of this mass to the central light output is more than 100 times the 
solar value. No normal population of stars has such a high mass-to-light ratio. 
This is consistent with the presence of a BH, but it does not rule out some 
other concentration of underluminous matter. 

Fortunatelly, the masers offer the possibility to have more convincing arguments. The Seyfert galaxy NGC~4258  
is one of the few nearby AGN known to possess nuclear water masers (the microwave 
equivalent of lasers). The enormous surface brightnesses ($\ge 10^{12}$ K), 
small sizes ($\le 10^{14}$ cm), and narrow linewidths (a few km s$^{-1}$) 
of these masers make them ideal probes of the structure and dynamics of the 
molecular gas in which they reside.  

VLBI observations (angular resolution 100 times better than that of HST)  
of the NGC~4258 maser have provided the first direct images of an AGN 
accretion disc, revealing a thin, subparsec-scale, differentially rotating 
warped disc in the nucleus of this relatively weak Seyfert~2 AGN.
The measurements imply that $4\times10^7$M$_{\odot}$ lie within half a 
lightyear of the centre (Herrnstein et al. 1999). This material
can hardly be a cluster of dead stars. In the case there were failed stars 
(brown dwarfs) that
never get hot enough to ignite the nuclear reactions that power stars because
their low mass ($<0.08$M$_{\odot}$), there would have to be many of them to 
explain the dark mass in NGC~4258. This would imply they would have to live 
very close together and most of them would collide with other brown dwarfs and 
the dark cluster would light up. 
In the case there were dead stars (WDs stars, NSs, or stellar mass BHs),  
these are more massive than brown dwarfs, so there would be fewer of them. It 
is well known that during the gravitational evolution of clusters of stars,
individual stars get ejected from the cluster, the remaining cluster contracts, and 
the evolution speeds up. Calculations show that a cluster of dead stars in 
NGC~4258 would evaporate completely in about 100 million years. This time is 
much less than the age of the galaxy. Therefore the most astrophysically plausible 
alternatives to a BH can be excluded. It seems unavoidable the 
conclusion that NGC~4258 contains a super-massive BH.      


\subsection{The BH in the Galactic centre}

The centre of our Galaxy is only 25000 lightyears away. Although its visible 
light is completely absorbed by intervening dust, its IR light penetrates the 
dust. The super-massive BH in the centre of the Milky Way was discovered as a 
bright non-thermal radio source in the 1970s and termed Sagittarius A$^*$ 
(Sgr A$^*$). The radio emission of SgrA$^*$ only varies slowly on time scales of several 
days to a few hundred days and generally with an amplitude $<10\%$. 
Potential X-ray radiation by Sgr A$^*$ was detected with the X-ray observatory 
ROSAT in the 1990s. A reliable identification of X-rays from Sgr A$^*$ was 
finally possible with the new X-ray satellites Chandra and XMM. 
A general review of Sgr A$^*$ can be found in Melia \& Falcke (2001).

\subsubsection{Motions of individual stars}

Two groups, Genzel (Munich) and Ghez (UCLA), have measured the motions 
of individual stars near the GC as projected on the plane of the sky. 
They used the speckle image technique.

\begin{figure}  
\begin{center}
\hspace{0.25cm}
   \psfig{figure=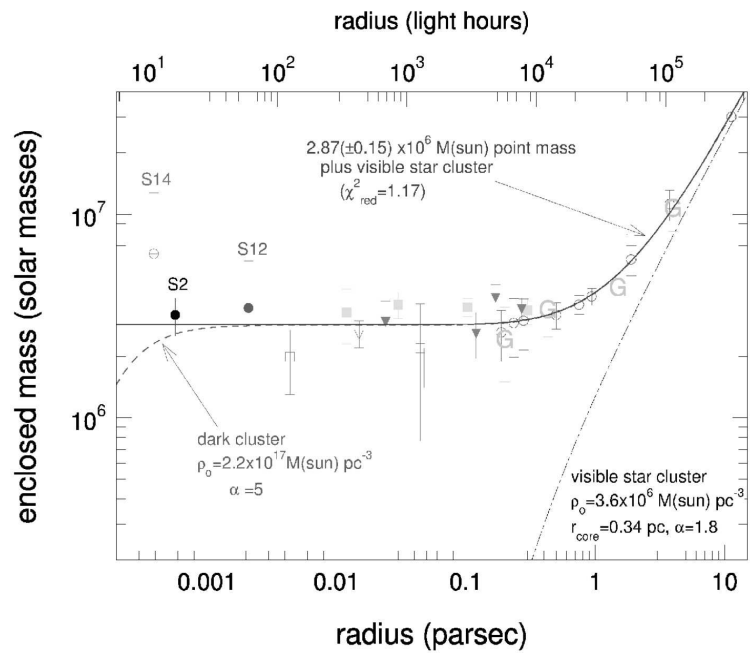,width=10.cm}
\caption{Central mass distribution in our Galaxy implied by the observed 
velocities measured  on the parsec and subparsec scale. The solid curve 
represents the stars plus a point mass  of $3.2\times 10^6$M$_{\odot}$. The dashed 
curve gives the contribution from the star cluster on the  parsec-scale. Figure reproduced from Sch\"odel et al. (2003).}
\label{massdis}
\end{center}
\end{figure}

The rapid motions show that there is a mass of $3\times10^6$M$_{\odot}$ centered 
on Sgr A$^*$. The mass enclosed inside a particular distance from Sgr A$^*$ 
stops dropping toward the centre at a distance of about 1 pc 
(see Figure~\ref{massdis}). This means that 
the mass in stars inside 3 lightyears has become negligible compared to the 
dark mass at the centre (Sch\"odel et al. 2003). 
As in NGC~4258, the implied density of matter is too high to allow a cluster 
of dark stars or stellar remnants.
NGC~4258 and the Milky Way give us important proof that BH exist.

\begin{figure}  
\begin{center}
\hspace{0.25cm}
   \psfig{figure=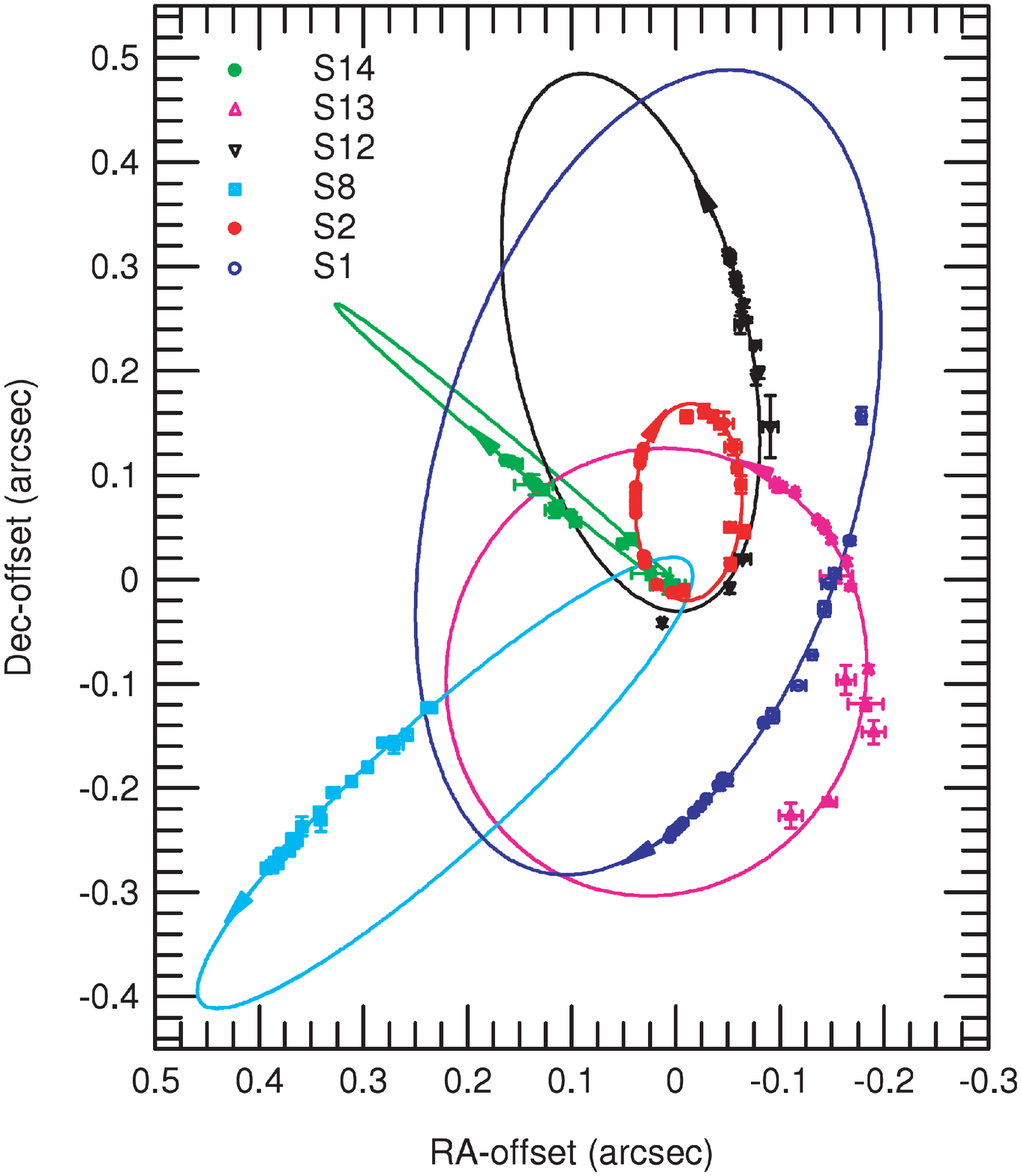,width=10.cm}
\caption{Projection on the sky of the six S stars included in 
the fitting. Figure reproduced from Eisenhauer et al. (2005).}
\label{orbits}
\end{center}
\end{figure}

The velocity dispersion increases to 400 km~s$^{-1}$ at a distance of 0.03 lightyears 
from Sgr A$^*$. Here stars have such small orbits that they revolve around 
the Galactic centre in a few decades. Motions in the plane of the sky have 
been measured for these stars for several  years (Eisenhauer et al. 2005;  Ghez et al. 2005). From a global fit to the positions and radial 
velocities of the six best stars (see Figure~\ref{orbits}) within 0.5"(=0.02 pc) of 
Sgr A$^*$ 
improved three dimensional stellar orbits were derived and their orbital parameters updated.
The updated estimate for the distance to the Galactic centre from the S2 
orbit fit is $7.62 \pm 0.32$ kpc, resulting in a central mass value of 
$(3.61 \pm 0.32) \times 10^6$M$_{\odot}$ (Eisenhauer et al. 2005).

\subsubsection{Size of Sgr A$^*$}

VLBI observations of Sgr A$^*$ have revealed an east--west elongated
structure whose apparent angular size at longer wavelengths is dominated by
the interstellar scattering angle, $\theta_{\rm obs}=\theta_{\rm obs}^{1cm}\lambda^2$, where $\theta_{\rm obs}$ is the observed size, in mas, at wavelength $\lambda$ in cm.
Thus, VLBI observations at shorter millimetre wavelengths, where the intrinsic
structure of Sgr A$^*$ could become comparable to the pure scattering size, are
expected to show deviations of the observed size from the scattering law. This has been demonstrated by the detection of the intrinsic size at 7\,mm (Bower et al. 2004). More recently, Zhi-Qiang Shen et al. (2005), carried out observations with the VLBA at 3.5\,mm of Sgr A$^*$.  They report a radio image of Sgr A$^*$, demonstrating that its size is 1\,AU. This high-resolution image of Sgr A$^*$, made at 3.5\,mm, also exhibits
an elongated structure. When combined with the lower limit on its mass, the lower limit on the mass
density is $6.5\times 10^{21}$M$_{\odot}$ pc$^{-3}$, which provides strong evidence that Sgr A$^*$ is a
super-massive BH.

\subsubsection{Near-infrared flares from Sgr A$^*$}

Detection of variable infrared (3.8 micron) light coincident to within 18 mas (1 $\sigma$) of the super-massive BH at the centre of the Milky Way galaxy and the radio source Sgr A$^*$ was reported by Ghez et al. (2004). The brightness variations, which occur on timescales as short as 40 minutes, reveal that the emission arises quite close to the BH, within 5 AU, or 80\,$r_{\rm Sch}$.
Two K-band flares observed on the 15th and 16th of June 2003 show a quasi-periodicity of about 17 min (Genzel et al. 2003). The only reasonable explanation for this so short period is that the oscillations are produced by Doppler boosting of hot gas near the last
stable orbit of a spinning Kerr BH.
The spin of the BH will allow for a last stable orbit closer to the event horizon and thus with a shorter orbital
frequency. From the observed 17 min quasi-periodicity it is estimated that the super-massive BH Sgr A$^*$ has a
spin that is half of the maximum possible spin of such an object (Genzel et al. 2003). Additional observations of flares and
their quasi-periodicity will be needed in order to confirm this result. 

\subsubsection{X-ray flares from Sgr A$^*$}

In the X-ray regime, SgrA$^*$ was found to exhibit two different states.
In the quiescent state, weak X-ray emission appears to come from a slightly
extended area around the BH that appears to be evidence of hot accreting
gas in the environment of SgrA$^*$.
SgrA$^*$ shows X-ray flares with a recurrence of about one per day. During these flares, the
emission rises by factors up to 100 during several tens of minutes and a distinctive
point source becomes visible at the location of SgrA$^*$. The short rise-and-decay
times of the flares suggest that the radiation must originate from a region within less
than 10\,$r_{\rm Sch}$ of a $2.6\times10^6$\,M$_{\odot}$ BH (Boganoff et al. 2001).

\subsubsection{Lensing of orbital shapes}

The frame--dragging that causes matter and light to co--rotate with the space--time
has effects in the orbits around BHs. Mueller (2007) shows  
relativistically distorted images of
tight orbits around BHs for different inclination angles and for Schwarzchild and Kerr BHs. When viewed by a distant observer at infinity, in nearly edge--on situations the observer would detect different apparent orbit
shapes depending on BH spin. These orbit shapes become asymmetric in the case of rotating BHs. The maximum relative spatial offset predicted is 0.08\,$r_{\rm g}=0.04r_{\rm Sch}$. This means that it might be possible to detect it by observing the Galactic centre BH but not by observing an stellar BH. The reason is that the angular size corresponding to an Schwarzschild radius of a super-massive BH is about three orders of magnitude larger than the an stellar mass BH. 
The angular size corresponding to an Schwarzschild radius is given by:

\begin{equation}
\theta_{\rm Sch}=2\,{\rm arctan}\,(r_{\rm Sch}/d) \simeq 2\,r_{\rm Sch}/d \simeq 39.4\times\Big(\frac{M}{10^6{\rm M}_{\odot}}\Big)\times\Big(\frac{1 {\rm kpc}}{d}\Big) \mu{\rm as}
\end{equation}

For a mass $M= 3.6\times 10^6\,{\rm M}_{\odot}$ at a distance $d=7.6\,$\,kpc one obtains $\theta_{\rm BH}=18.6\,\mu{\rm as}$. Then, the angular size of the relative spatial offset predicted is $0.7\,\mu{\rm as}$.

In contrast, for a close stellar BH like Cygnus X$-$1, $M$=10\,${\rm M}_{\odot}$ 
and $d$=2.2\,kpc, the angular size is $\theta_{\rm Sch}\simeq 10^{-3}\mu{\rm as}$, which is an extremely low value for the current instrumentation.



\acknowledgments I would like to thank G.E. Romero and P. Benaglia, for their invitation and the excellent organisation of the International School.
I am indebted to J. Mart\'{\i} and V. Zabalza for a careful reading of the manuscript and their valuable comments.
The author acknowledges support of the Spanish Ministerio de Educaci\'on y Ciencia (MEC) under grant AYA2007-68034-C03-01 and FEDER funds.

\end{document}